\pdfoutput=1
\documentclass[english,english,prl,english,preprint,amsmath,amssymb,aps,longbibliography,showkeys, titlepage]{revtex4-2}
\usepackage{lmodern}
\usepackage[T1]{fontenc}
\usepackage[latin9]{inputenc}
\usepackage{amsmath}
\usepackage{amsthm}
\usepackage{graphicx}

\makeatletter
\usepackage[T1]{fontenc}
\usepackage{amsmath}
\usepackage{amsthm}
\usepackage{amssymb}
\usepackage{graphicx}

\usepackage{graphicx}
\usepackage{float}
\usepackage{xcolor}
\usepackage[hypertexnames=false]{hyperref}
\hypersetup{
	breaklinks = true,
    colorlinks = true,
    citecolor = {blue},
	urlcolor = {blue},
	linkcolor = {blue}
}

\makeatother

\usepackage{babel}
\begin{document}
\title{Oblique photons, plasmons, and current-plasmons in relativistic plasmas
and their topological implications}
\author{Hong Qin}
\email{hongqin@princeton.edu }

\affiliation{Princeton Plasma Physics Laboratory, Princeton University, Princeton,
NJ 08540}
\affiliation{Department of Astrophysical Sciences, Princeton University, Princeton,
NJ 08540}
\author{Eric Palmerduca }
\email{ep11@princeton.edu }

\affiliation{Princeton Plasma Physics Laboratory, Princeton University, Princeton,
NJ 08540}
\affiliation{Department of Astrophysical Sciences, Princeton University, Princeton,
NJ 08540}
\begin{abstract}
Photons in vacuum are transverse in any inertial frame; longitudinal
photons only exist virtually. By developing a manifestly covariant
theory for electromagnetic excitations in relativistic plasmas and
applying Wigner's little group method for elementary particle classifications,
we show that photons in plasmas are neither transverse nor longitudinal;
they are oblique. Plasmons are electromagnetic and oblique as well.
The Lorentz invariant characteristics that distinguishes photons and
plasmons is covariant compressibility. The manifestly covariant theory
predicts the existence of the current-plasmon, a third oblique, electromagnetic
eigenmode, and it also enables the study of photon topology in plasmas.
Plasmas remove the photon's Dirac point in vacuum by giving it an
effective mass, but create a tilted Dirac-Weyl point by reviving the
virtual longitudinal photon. The manifest covariance of the theory
demonstrates that relativistic transparency, despite being widely
studied, does not exist in plasmas.
\end{abstract}
\maketitle
Photons in vacuum are transverse in any inertial frame; longitudinal
photons only exist virtually. But we show that in plasmas, media with
freely moving charged particles, the characterization of transversality
and longitudinality for the electromagnetic excitations are not covariant.
In plasmas, photons are neither transverse nor longitudinal; they
are oblique. Plasmons, which correspond to the virtual longitudinal
photons in vacuum, are oblique and electromagnetic as well. 

Only in the special inertial frame co-moving with the plasma, photons
are transverse and plasmons are longitudinal and electrostatic. However,
albeit non-covariant, this is how the classical theory of photons
and plasmons in plasmas is developed. The shortcomings of this classical,
non-covariant approach are well-known. For example, the plasmon, a.k.a.
the electrostatic Langmuir wave, will carry a time-dependent magnetic
component in a general inertial frame, where the electrostatic assumption
in the non-covariant theory breaks down. Nevertheless, for an electrostatic
plasmon observed in a rest frame, it is still important to know what
kind of electromagnetic excitation it would be and whether it is a
well-defined eigenmode in a moving inertial frame. 

To address these questions, we develop a systematic, manifestly covariant
theory for electromagnetic excitations in plasmas. To focus on the
main interests of the present study, the photons and plasmons, we
take the minimal theoretical model---an unmagnetized, homogeneous,
cold plasma. For the theory to be covariant, the plasma carries a
general 4-velocity $u_{0}=\left(\gamma,\gamma\beta_{1},\gamma\beta_{2},\gamma\beta_{3}\right)^{T}$
for all charged species, and no a priori assumptions about the polarizations
of the modes are allowed. For instance, we should not assume any mode
is transverse, longitudinal, or electrostatic. In fact, as we will
show, none of the covariant modes are. We don't even know a priori
if the covariant theory admits eigenmodes that can be identified as
photons or plasmons as described by the standard non-covariant theory. 

Previous work has considered the electromagnetic waves and electrostatic
waves in relativistic plasmas \citep{Akhiezer1955,Kaw1970,Goloviznin1999,Goloviznin2000,Cattani2000,Mahajan2017}.
However, despite our best effort, we have found no systematic, manifestly
covariant treatment of the subject in the existing literature. Most
previous studies focused on relativistic corrections or effects on
specific modes, and dispersion relations of the modes were often written
down by heuristic arguments based on the requirement of covariance
or the necessity of the relativistic factor $\gamma$ and with a priori
assumptions about the dispersion relations and polarization properties.
Because of the lacking of a systematic, manifestly covariant treatment,
certain important phenomena, such as the self-induced transparency
in relativistic plasmas, have been debated about in the literature
\citep{Akhiezer1955,Kaw1970,Lefebvre1995,Guerin1996,Goloviznin1999,Goloviznin2000,Cattani2000,Eremin2010,Mahajan2017}.
With the arrival of increasingly powerful laser facilities, we hope
the results reported in the present study contribute to the understanding
on this subject.

The main obstacle to overcome for a manifestly covariant theory is
how to model the collective electromagnetic excitations in plasmas
using covariant fields. The standard governing equations expressed
in terms of the electromagnetic fields $\boldsymbol{E}$ and $\boldsymbol{B}$,
plasma velocity $\boldsymbol{v}$, and plasma density $n$ are not
manifestly covariant, because $\boldsymbol{E}$, $\boldsymbol{B}$,
and $\boldsymbol{v}$ are 3D vectors instead of covariant vectors
or forms in the 4D spacetime. For a manifestly covariant theory, it
is necessary to use exclusively the 4-potentional 1-form $A$ or the
electromagnetic tensor $F\equiv dA$ and the 4-velocity $u$. The
difficulties associated with using $A$ and $u$ are two-fold. The
unfixed gauge freedom may introduce nonphysical modes that need to
be discarded, and the on-shell condition of $u$ imposes a constraint
on allowed eigenmodes in the classical regime. For the present study,
it happens that the nonphysical gauge modes and off-shell virtual
modes can be easily identified and removed.

Another benefit of the manifestly covariant theory is that symmetry
group methods can be applied. We will apply Wigner's little group
method for the classification of elementary particles to classify
the electromagnetic excitations in plasmas. 

Using the manifestly covariant theory and classification techniques
developed, we rigorously show that the system admits one 2D mode that
can be identified as a photon and one 1D mode that can be identified
as a plasmon, as well as their anti-particles. But the photons are
neither transverse nor longitudinal. The electric field of photons
is oblique to its momentum. Plasmons are not electrostatic, and they
are also oblique. In addition, the manifestly covariant theory also
predicts the existence of the current-plasmon, which is a third, previously
unknown oblique electromagnetic excitation in relativistic plasmas.
If photons, plasmons, and current-plasmons all are electromagnetic
and oblique in general inertial frames, is there a covariant label
that can be used to distinguish them? We find that covariant compressibility
$\chi$ is such a label. Photons and current-plasmons are not covariantly
compressible, but plasmons are. The manifestly covariant theory also
enables the study of photon topology in plasmas. We will show that
plasmas remove the photon's upright Dirac point in vacuum by giving
it an effective mass, but create a tilted Dirac-Weyl point by reviving
the virtual longitudinal photon. In addition, the manifest covariance
of the theory demonstrates that relativistic transparency \citep{Akhiezer1955,Kaw1970,Lefebvre1995,Guerin1996,Goloviznin1999,Goloviznin2000,Cattani2000,Eremin2010,Mahajan2017},
despite being widely studied, does not exist in plasmas.

Our starting point is the following manifestly covariant system governing
the electromagnetic excitations in an unmagnetized cold plasma,
\begin{align}
\partial_{\mu}J_{s}^{\mu} & =0,\label{eq:pJ}\\
\partial_{\mu}F^{\mu\nu} & =-4\pi\sum_{s}J_{s}^{\nu},\label{eq:pF}\\
u_{s\mu}F^{\mu\nu} & =-\frac{m_{s}c^{2}}{q_{s}}u_{s\mu}\partial^{\mu}u_{s}^{\nu},\label{eq:uF}
\end{align}
where subscript ``$s$'' denotes particle species, $q_{s}$ and
$m_{s}$ are the charge and mass of species $s$ in the rest frame,
$u_{s}^{\mu}=\left(\gamma_{s},\gamma_{s}v_{s}^{i}/c\right)$ is the
4-velocity of the species $s$, $n_{s}$ is its density in the momentarily
co-moving reference frame (MCRF) \citep{Schutz2009}, and $J_{s}^{\mu}\equiv q_{s}n_{s}u_{s}^{\mu}$
is the 4-current. The sign and index conventions are the same as Misner,
Thorn, and Wheeler \citep{Misner1973}. 

The electromagnetic tensor $F$ is defined by the 1-form potential
as $F\equiv dA.$ We will take the Lorenz gauge $\partial_{\mu}A^{\mu}=0$,
which is Lorentz invariant, i.e., covariant. In terms of $A,$ Eqs.\,(\ref{eq:pF})
and (\ref{eq:uF}) are, respectively, 
\begin{align}
\partial_{\mu}\partial^{\mu}A^{\nu} & =-4\pi\sum_{s}J_{s}^{\nu},\label{eq:ppA}\\
2u_{s\mu}\partial^{[\mu}A^{\nu]} & =-\frac{m_{s}c^{2}}{q_{s}}u_{s\mu}\partial^{\mu}u_{s}^{\nu}.\label{uA}
\end{align}
The unperturbed system is assumed to be uniform over spacetime and
has a vanishing 4-potential, i.e., $A_{0}=0.$ Each charged species
has a constant equilibrium 4-velocity $u_{s0}$ and a constant equilibrium
density $n_{s0}.$ The unperturbed system does not carry net 4-current.
To simplify the problem to the bare minimum for our present interest,
it is assumed that the equilibrium 4-velocity is the same for all
species, 
\[
u_{s0}=u_{0}=\left(\gamma,\gamma\beta_{1},\gamma\beta_{2},\gamma\beta_{3}\right)^{T}.
\]
The electromagnetic excitation is considered to be a small perturbation
of the system. We further assume that the ions are heavy and only
electrons are perturbed, i.e., 
\begin{align}
n_{e} & =n_{e0}+\delta n,\\
u_{e} & =u_{0}+\delta u,\\
A & =\delta A.
\end{align}
The linearized system governing the dynamics of $\delta n$, $\delta u$,
and $\delta A$ is 
\begin{align}
u_{0}^{\mu}\partial_{\mu}\delta n+n_{0}\partial_{\mu}\delta u^{\mu} & =0,\\
2u_{0\mu}\partial^{[\mu}\delta A^{\nu]}+\frac{m_{e}c^{2}}{q_{e}}u_{0\mu}\partial^{\mu}\delta u^{\nu} & =0,\\
\partial_{\mu}\partial^{\mu}\delta A^{\nu}+4\pi q_{e}\left(\delta nu_{0}^{\nu}+n_{e0}\delta u^{\nu}\right) & =0.
\end{align}
Let $\delta A^{\mu},$ $\delta u^{\mu},$ $\delta n$ $\sim\exp\left(ik_{\mu}x^{\mu}\right),$
where $k_{\mu}=(-\omega,k_{1},k_{2},k_{3}).$ The system reduces to
\begin{gather}
M\left(k,u_{0},\omega_{p}\right)\psi=0,\label{eq:Mpsi}\\
\psi\equiv\left(\delta A^{\mu},\delta u^{\mu},\delta n\right)^{T},\\
M\left(k,u_{0},\omega_{p}\right)\equiv\left(\begin{array}{ccc}
-k_{\mu}k^{\mu}I_{4\times4} & \omega_{p}^{2}I_{4\times4} & \omega_{p}^{2}u_{0}^{\mu}\\
u_{0\mu}k^{\mu}I_{4\times4}-k^{\mu}u_{0\nu} & u_{0\mu}k^{\mu}I_{4\times4} & 0\\
0 & k_{\mu} & u_{0}^{\mu}k_{\mu}
\end{array}\right),
\end{gather}
where $M\left(k,u_{0},\omega_{p}\right)$ is a $9\times9$ matrix
and $\psi$ is the perturbed fields $\delta A^{\mu},$ $\delta u^{\mu},$
and $\delta n$ arranged into a $9\times1$ column vector. We will
also denote $\psi$ by 
\begin{equation}
\left\{ \left(\delta A^{0},\delta A^{1},\delta A^{2},\delta A^{3}\right),\left(\delta u^{0},\delta u^{1},\delta u^{2},\delta u^{3}\right),\delta n\right\} 
\end{equation}
for easy identification of its components. The superscript ``$\mu$''
in $\delta A^{\mu}$ and $\delta u^{\mu}$ indicates that they are
the contravariant components of the perturbed field. In addition,
$\delta A$ has been normalized by $m_{e}c^{2}/q_{e}$ and $\delta n$
by $n_{e0}$, and $\omega_{p}^{2}\equiv4\pi n_{e0}q_{e}^{2}/m_{e}c^{3}$
is the normalized plasma frequency-squared. Note that $n_{e0}$ and
$m_{e}$ are quantities measured in the MCRF. The dispersion relation
of the eigenmodes is thus
\begin{equation}
D\left(k,u_{0},\omega_{p}\right)\equiv\det\left(M\left(k,u_{0},\omega_{p}\right)\right)=0.
\end{equation}
For a given $\left(u_{0},\omega_{p}\right),$ each root of $D\left(k,u_{0},\omega_{p}\right)$
is an eigen 4-wavevector of the system, and the kernel (null space)
of $M$ at each eigen 4-wavevector $k$ is the corresponding eigenvector
space. Note that not all vectors in the kernel of $M$ at a given
eigen 4-wavevector $k$ are physically valid eigenvectors. Two constraints
need to be satisfied, the Lorenz gauge condition,
\begin{equation}
k_{\mu}\delta A^{\mu}=0,\label{eq:kpA}
\end{equation}
and the linearized on-shell condition (normalization condition),
\begin{equation}
u_{0\mu}\delta u^{\mu}=0.\label{eq:upu}
\end{equation}

The eigen 4-wavevectors $k$ and eigenvectors $\psi$ of the system
determined by Eq.\,(\ref{eq:Mpsi}) are algebraically complicated.
Direct calculations are formidable. However, the system can be significantly
simplified by choosing a specific inertial frame. This is enabled
by the fact that the system is manifestly covariant, which can described
as follows. For a Lorentz transformation 
\begin{equation}
x^{\nu}\rightarrow x^{\prime\mu}=\Lambda_{\ \nu}^{\mu}x^{\nu},
\end{equation}
specified by a Lorentz matrix $\Lambda_{\ \nu}^{\mu},$ the 4-vector
fields of $u_{0},$ $k,$ $\delta u$, and $\delta A$ in the original
frame $x$ transform into $u_{0}^{\prime},$ $k^{\prime},$ $\delta u^{\prime}$,
and $\delta A^{\prime}$ as contravariant 4-vectors in the frame $x^{\prime}.$
The density field $\delta n$ is a scalar because it is always measured
in the MCRF. It is straightforward to verify that 
\begin{align}
\Lambda_{9} & M\left(k,u_{0},\omega_{p}\right)\psi=M\left(k^{\prime},u_{0}^{\prime},\omega_{p}\right)\psi^{\prime},\label{eq:LamdaM}\\
\psi^{\prime} & \equiv\Lambda_{9}\psi,\\
\Lambda_{9} & \equiv\left(\begin{array}{ccc}
\Lambda_{\ \nu}^{\mu}\\
 & \Lambda_{\ \nu}^{\mu}\\
 &  & 1
\end{array}\right).
\end{align}
We will also use the following shorthand notations,
\begin{align}
M & \equiv M\left(k,u_{0},\omega_{p}\right),\thinspace\thinspace\thinspace M^{\prime}\equiv M\left(k^{\prime},u_{0}^{\prime},\omega_{p}\right),\\
D & \equiv\det\left(M\right),\thinspace\thinspace\thinspace D^{\prime}\equiv\det\left(M^{\prime}\right).
\end{align}
Equation (\ref{eq:LamdaM}) implies that $M$ and $M^{\prime}$ are
similar, i.e.,
\begin{equation}
\Lambda_{9}M=M^{\prime}\Lambda_{9},
\end{equation}
and thus 
\begin{equation}
D=D^{\prime}.
\end{equation}
For each eigen 4-wavevector $k$ of $D,$ $k^{\prime}\equiv\Lambda k$
is an eigen 4-wavevector of $D^{\prime},$ and for each vector $\psi$
in the kernel of $M$ at $k,$ $\psi^{\prime}\equiv\Lambda_{9}\psi$
is a vector in the kernel of $M^{\prime}$ at $k^{\prime}$. The isomorphism
between $\left(k^{\prime},\psi^{\prime},M^{\prime}\right)$ and $\left(k,\psi,M\right)$
reflects the fact that eigenmodes $\left(k^{\prime},\psi^{\prime}\right)$
and $\left(k,\psi\right)$ are actually one identical physical mode
seen by two different observers in the frame $x^{\prime}$ and the
frame $x$, respectively. With this isomorphism, we can label the
physical modes in the frame $x$ by their representation in a specially
chosen frame $x^{\prime}$ where the eigen 4-wavevectors $k^{\prime}$
are easier to solve for. It is worthwhile to emphasize that such an
isomorphism is not only for the purpose of calculation and classification
of eigenmodes, but also a necessity required by the covariance. Eigenmodes
in different frames linked by the isomorphism must be treated as one
identical mode according to the laws of relativity. As often is the
case, preserving the symmetry required by the physics simplifies the
seemingly complex calculations. 

It is evident that in a frame boosted with a 4-velocity equal to the
equilibrium 4-velocity $u_{0}$, the equilibrium plasma will be at
rest, i.e., 
\begin{equation}
u_{0}^{\prime}=(1,0,0,0)^{T}.
\end{equation}
 The Lorentz matrix for such a boost is 
\begin{equation}
\Lambda\left(\beta_{1},\beta_{2},\beta_{3}\right)=\left(\begin{array}{cccc}
\gamma & -\gamma\beta_{1} & -\gamma\beta_{2} & -\gamma\beta_{3}\\
-\gamma\beta_{1} & 1+\left(\gamma-1\right)\frac{\beta_{1}^{2}}{\beta^{2}} & \left(\gamma-1\right)\frac{\beta_{1}\beta_{2}}{\beta^{2}} & \left(\gamma-1\right)\frac{\beta_{1}\beta_{3}}{\beta^{2}}\\
-\gamma\beta_{2} & \left(\gamma-1\right)\frac{\beta_{1}\beta_{2}}{\beta^{2}} & 1+\left(\gamma-1\right)\frac{\beta_{2}^{2}}{\beta^{2}} & \left(\gamma-1\right)\frac{\beta_{2}\beta_{3}}{\beta^{2}}\\
-\gamma\beta_{3} & \left(\gamma-1\right)\frac{\beta_{1}\beta_{3}}{\beta^{2}} & \left(\gamma-1\right)\frac{\beta_{2}\beta_{3}}{\beta^{2}} & 1+\left(\gamma-1\right)\frac{\beta_{3}^{2}}{\beta^{2}}
\end{array}\right),\label{eq:Lamda}
\end{equation}
where $\beta^{2}\equiv\sum_{j}\beta_{j}^{2}$. We will also refer
to the original frame $x$ as the lab frame, where the plasma has
a nontrivial equilibrium 4-velocity $u_{0}$. The frame $x^{\prime}$
will be referred to as a rest frame for the obvious reason.

We now further transform the inertial frame $x^{\prime}$ by an additional
spatial rotation, i.e., 
\begin{equation}
x^{\prime\nu}\rightarrow x^{\prime\prime\mu}=\Lambda_{\ \nu}^{\prime\mu}x^{\prime\nu},
\end{equation}
where $\Lambda^{\prime}\in\text{SO}(3)\subset\text{SO}^{+}(1,3)$
does not change the 0-th component of $x^{\prime}.$ We choose $\Lambda^{\prime}$
such that in the frame $x^{\prime\prime}$, the spatial component
of $k^{\prime\prime}$ is in the 1st-direction, 
\begin{equation}
k^{\prime\prime}=\left(\Omega,K,0,0\right)^{T}.
\end{equation}
Here, we use $\Omega$ and $K$ to denote the mode frequency and wavenumber
in the frame $x^{\prime\prime}.$ Because $\Lambda^{\prime}\in\text{SO}(3)$,
we have
\begin{align}
u_{0}^{\prime\prime} & =u_{0}^{\prime}=\left(1,0,0,0\right)^{T},\label{eq:upp}\\
\Omega & =\omega^{\prime}=\Lambda_{\ \mu}^{0}k^{\mu},\\
K^{2} & =k_{j}^{\prime}k^{\prime j}=k_{\nu}\Lambda_{\ j}^{\nu}\Lambda_{\ \mu}^{j}k^{\mu}.
\end{align}
As expected, Eq.\,(\ref{eq:upp}) says that $x^{\prime\prime}$ is
also a rest frame. In the frame $x^{\prime\prime},$ the matrix $M^{\prime\prime}\equiv M\left(k^{\prime\prime},u_{0}^{\prime\prime},\omega_{p}\right)$
is simplified to
\begin{equation}
M^{\prime\prime}=\left(\begin{array}{ccccccccc}
\Omega^{2}-K^{2} & 0 & 0 & 0 & \omega_{p}^{2} & 0 & 0 & 0 & \omega_{p}^{2}\\
0 & \Omega^{2}-K^{2} & 0 & 0 & 0 & \omega_{p}^{2} & 0 & 0 & 0\\
0 & 0 & \Omega^{2}-K^{2} & 0 & 0 & 0 & \omega_{p}^{2} & 0 & 0\\
0 & 0 & 0 & \Omega^{2}-K^{2} & 0 & 0 & 0 & \omega_{p}^{2} & 0\\
0 & 0 & 0 & 0 & -\Omega & 0 & 0 & 0 & 0\\
K & -\Omega & 0 & 0 & 0 & -\Omega & 0 & 0 & 0\\
0 & 0 & -\Omega & 0 & 0 & 0 & -\Omega & 0 & 0\\
0 & 0 & 0 & -\Omega & 0 & 0 & 0 & -\Omega & 0\\
0 & 0 & 0 & 0 & -\Omega & K & 0 & 0 & -\Omega
\end{array}\right).
\end{equation}
It is straightforward to calculate the dispersion relation in the
frame $x^{\prime\prime}$ from $M^{\prime\prime}$,
\begin{equation}
D^{\prime\prime}\equiv\det\left(M^{\prime\prime}\right)=-\Omega^{3}\left(\Omega^{2}-K^{2}\right)^{2}\left(\Omega^{2}-\omega_{p}^{2}\right)\left(\Omega^{2}-K^{2}-\omega_{p}^{2}\right)^{2}.
\end{equation}
 Detailed calculations show that the spectrum of $M^{\prime\prime}$
consists of the following 5 classes. 
\begin{enumerate}
\item Photons and anti-photons. For the eigenmodes at
\begin{equation}
\Omega=\sqrt{\omega_{p}^{2}+K^{2}}\ \ \text{and}\ \ \Omega=-\sqrt{\omega_{p}^{2}+K^{2}},\label{eq:Photonx''}
\end{equation}
the corresponding eigenvector space for the modes is
\begin{equation}
\left\{ \left(0,0,0,-1\right),\left(0,0,0,1\right),0\right\} \oplus\left\{ \left(0,0,-1,0\right),\left(0,0,1,0\right),0\right\} ,
\end{equation}
which is 2-dimensional. There is no density perturbation, and in this
co-moving frame, the modes are transverse, i.e., 
\begin{equation}
\delta\boldsymbol{E}^{\prime\prime}\cdot\boldsymbol{k}^{\prime\prime}=\delta E_{j}^{\prime\prime}k^{\prime\prime j}=0.
\end{equation}
Because of these polarization characteristics, we identify the mode
at $\Omega=\sqrt{\omega_{p}^{2}+K^{2}}$ as the photon and the mode
at $\Omega=-\sqrt{\omega_{p}^{2}+K^{2}}$ as the anti-photon.
\item Plasmons and anti-plasmons. There are two modes at
\begin{equation}
\Omega=\omega_{p}\ \ \text{and}\ \ \ensuremath{\Omega=-\omega_{p}.}
\end{equation}
 The eigenvector space for the $\Omega=\omega_{p}$ mode is 
\begin{equation}
\left\{ \frac{\omega_{p}^{2}}{K^{2}-\omega_{p}^{2}}\left(1,\frac{\omega_{p}}{K},0,0\right),\left(0,\frac{\omega_{p}}{K},0,0\right),1\right\} ,
\end{equation}
and that for the $\Omega=-\omega_{p}$ mode is 
\begin{equation}
\left\{ \frac{\omega_{p}^{2}}{K^{2}-\omega_{p}^{2}}\left(1,-\frac{\omega_{p}}{K},0,0\right),\left(0,-\frac{\omega_{p}}{K},0,0\right),1\right\} .
\end{equation}
In this co-moving frame, both modes are electrostatic and longitudinal,
i.e., 
\begin{align}
\delta\boldsymbol{B}^{\prime\prime} & =0,\\
\boldsymbol{k}^{\prime\prime}\times\delta\boldsymbol{E}^{\prime\prime}=\epsilon_{ijl}k^{\prime\prime j}\delta E^{\prime\prime l} & =0.
\end{align}
We identify the first mode as the plasmon and the other as the anti-plasmon.
\item Current-plasmons. Two of the eigenvectors at $\Omega=0$ are magnetostatic
perturbations generated by plasma current perturbation. Their eigenvector
space is 
\begin{equation}
\left\{ \left(0,0,0,\frac{\omega_{p}^{2}}{K^{2}}\right),\left(0,0,0,1\right),0\right\} \oplus\left\{ \left(0,0,\frac{\omega_{p}^{2}}{K^{2}},0\right),\left(0,0,1,0\right),0\right\} .
\end{equation}
The polarization directions of the modes are similar to that of photons,
but the phase between the current and vector potential is shifted
by $\pi.$ The relative amplitude between the magnetic field and normalized
flow velocity is proportional to the plasma density, akin to ratio
between electrical field and normalized density in the case of plasmons.
In this sense, this 2D mode is a hybrid between photons and plasmons.
We will call it current-plasmon. It was not identified in previous
studies, probably because it has a zero frequency in the rest frames
$x^{\prime}$ and $x^{\prime\prime}.$ But in the lab frame $x$,
the mode is electromagnetic, oblique, and propagating (see below).
\item Gauge modes. The eigenmodes at $\Omega=\pm K$ have the same dispersion
relations as the photon and anti-photon in vacuum. The eigenvectors
for $\Omega=K$ and $\Omega=-K$ are, respectively, 
\begin{equation}
\left\{ \left(1,1,0,0\right),\left(0,0,0,0\right),0\right\} \ \ \text{and\ }\ \left\{ \left(-1,1,0,0\right),\left(0,0,0,0\right),0\right\} .
\end{equation}
These modes carry no electric, magnetic, velocity, or density perturbations.
They are the gauge modes introduced by the unfixed gauge freedom,
and will be regarded as trivial modes that have no observable effects
in the classical regime. 
\item Off-shell virtual mode. Since $\Omega=0$ is a triple root of the
$D^{\prime\prime}=0,$ in addition to the 2D current-plasmon identified
above, the system admits another eigenvector at $\Omega=0$, which
is
\begin{equation}
\left\{ \left(0,0,0,0\right),\left(-1,0,0,0\right),1\right\} .
\end{equation}
However, it does not satisfy the on shell condition (\ref{eq:upu}).
It is not a valid physical mode in the classical regime and will be
discarded. 
\end{enumerate}
The classification of the electromagnetic excitations described above
resembles Wigner's little group method for the classification of massive
particles \citep{Wigner1939,Simms1968,Weinberg1995,Palmerduca2023},
except that the base manifold in the present study is the product
of the space of equilibrium flow $u_{0}$ and the space of 4-wavevector
$k$. The Lorentz transformation applies to $u_{0}$ and $k$ simultaneously. 

The three modes identified, i.e., photons, plasmons, and current plasmons,
were classified according to their properties in the specially chosen
co-moving frame $x^{\prime\prime}$. But the labels for the classification,
i.e., transversality, longitudinality, electrostaticity, and magnetostaticity,
are not covariant. In the lab frame $x$, the polarizations and dispersion
relations are expected to be very different. Because the theory is
manifestly covariant, the dispersion relations and eigenvectors in
the lab frame $x$ can be easily obtained via the Lorentz transformation,
\begin{align}
\left(\Omega,K,0,0\right)^{T} & =k^{\prime\prime}=\Lambda^{\prime}\Lambda k,\label{eq:kpp}\\
\psi & =\Lambda^{-1}\Lambda^{\prime-1}\psi^{\prime\prime}.\label{eq:psipp}
\end{align}
Specifically, pushing forward the 4-wavevector $k$ in the lab frame
$x$, i.e., expressing the dispersion relation in the frame $x^{\prime\prime}$
in terms of $k$ via Eq.\,(\ref{eq:kpp}), gives the dispersion relation
in the lab frame $x$. The pull-back of $\psi^{\prime\prime}$ in
the fame $x^{\prime\prime}$ according to Eq.\,(\ref{eq:psipp})
is the eigenvector $\psi$ in the lab frame $x$. We list below the
properties of photons, plasmons, and current-plasmons in the lab frame
$x$. 
\begin{enumerate}
\item Photons and anti-photons in the lab frame. Simple calculation according
to the push-forward algorithm given above shows that the dispersion
relations for photons and anti-photons in the lab frame $x$ are,
respectively,
\begin{equation}
\omega=\sqrt{\omega_{p}^{2}+k_{j}k^{j}}\ \ \text{and}\ \ \omega=-\sqrt{\omega_{p}^{2}+k_{j}k^{j}},\label{eq:photonDR}
\end{equation}
which is expected because $\omega^{2}-k_{j}k^{j}$ is a Lorentz invariant
scalar. The dispersion relations are identical to their counterparts,
Eq.\,(\ref{eq:Photonx''}), in the rest frame $x^{\prime\prime}.$
A direct consequence of this result is that the cutoff frequency for
photons in a relativistic plasma does not depend on how fast the plasma
is moving in the lab frame $x$. The manifest covariance of the dispersion
relation demonstrates that relativistic transparency \citep{Akhiezer1955,Kaw1970,Lefebvre1995,Guerin1996,Goloviznin1999,Goloviznin2000,Cattani2000,Eremin2010,Mahajan2017},
despite being widely studied, does not exist in plasmas. However,
the polarization properties of the eigenmodes change under the Lorentz
transformations. In the frame $x^{\prime},$ photons are still transverse
because $\Lambda^{\prime-1}\in\text{SO}(3),$ which rotates both $\delta\boldsymbol{E}^{\prime\prime}$
and $\boldsymbol{k}^{\prime\prime}$as 3D vectors and preserves the
transversality. On the other hand, $\Lambda^{-1}$ is a boost, and
it transforms $\boldsymbol{k}^{\prime}$ as the spatial components
of the contravariant 4-vector $k^{\prime}$ and $\delta\boldsymbol{E}^{\prime}$
as the temporal components of the electromagnetic tensor $F$. The
transversality is broken by a general boost $\Lambda^{-1}$ unless
$\omega_{p}$ vanishes. To see this, let's take the example of 
\begin{align}
\psi^{\prime} & =\left\{ \left(0,0,0,-1\right),\left(0,0,0,1\right),0\right\} ,\\
k^{\prime} & =(\sqrt{\omega_{p}^{2}+K^{2}},K,0,0),
\end{align}
which is a transverse photon in the rest frame $x^{\prime}$ propagating
along the 1st-direction with electrical field polarized in the $3$rd-direction.
Assume the plasma flow in the lab frame $x$ is in the 3rd-direction,
i.e., 
\begin{equation}
\Lambda^{-1}=\Lambda\left(0,0,-\beta\right),\label{eq:Lamda3}
\end{equation}
then in the lab frame $x$, we find
\begin{equation}
\cos\theta_{1}\equiv\frac{\left|\delta\boldsymbol{E}\cdot\boldsymbol{k}\right|}{\left|\delta\boldsymbol{E}\right|\left|\boldsymbol{k}\right|}=\frac{\gamma\left|\beta\right|}{\sqrt{\left(\frac{\omega^{2}}{\omega_{p}^{2}}-\beta^{2}\gamma^{2}\right)\left(\frac{\omega^{2}}{\omega_{p}^{2}}-1\right)}}\le1,\label{eq:theta1}
\end{equation}
where $\theta_{1}$ is the pitch angle between $\delta\boldsymbol{E}$
and $\boldsymbol{k}$. This expression shows that photons carry an
electrical field in the wavevector direction unless the plasma disappears
($\omega_{p}=0$) or is motionless \textbf{$(\beta=0)$}. The maximum
value of $\cos\theta_{1}$ is reached at $\omega/\omega_{p}=\gamma$,
which is the smallest value of $\omega/\omega_{p}$ under the boost
in the 3rd-direction. In general, photons and anti-photons in plasmas
are neither transverse nor longitudinal; they are oblique. 
\item Plasmons and anti-plasmons in the lab frame. By the same push-forward
algorithm above, the dispersion relations for plasmons and anti-plasmons
in the lab frame $x$ are found to be 
\begin{equation}
\omega=k^{j}\beta_{j}+\frac{\omega_{p}}{\gamma}\ \ \text{and}\ \ \omega=k^{j}\beta_{j}-\frac{\omega_{p}}{\gamma},
\end{equation}
which can be written in a manifestly covariant manner as
\begin{equation}
k^{\mu}u_{0\mu}=-\omega_{p}\ \ \text{and}\ \ k^{\mu}u_{0\mu}=\omega_{p}.
\end{equation}
Importantly, in the lab frame $x$, plasmons are electromagnetic and
oblique as photons. For example, take the electrostatic longitudinal
plasmon in the rest frame $x^{\prime}$ with
\begin{align}
\psi^{\prime} & =\left\{ \frac{\omega_{p}^{2}}{K^{2}-\omega_{p}^{2}}\left(1,\frac{\omega_{p}}{K},0,0\right),\left(0,\frac{\omega_{p}}{K},0,0\right),1\right\} ,\\
k^{\prime} & =(\omega_{p},K,0,0).
\end{align}
Assume again that the plasma flow is in the 3rd-direction in the lab
frame $x$, with the Lorentz matrix given by Eq.\,(\ref{eq:Lamda3}).
In the lab frame $x,$ the plasmon has a non-vanishing magnetic component
in the 2nd-direction, 
\[
\delta\boldsymbol{B}=(0,-i\beta\gamma,0)^{T},
\]
 and the cosine of the pitch angle $\theta_{2}$ between $\delta\boldsymbol{E}$
and $\boldsymbol{k}$ is 
\begin{equation}
\cos\theta_{2}\equiv\frac{\left|\delta\boldsymbol{E}\cdot\boldsymbol{k}\right|}{\left|\delta\boldsymbol{E}\right|\left|\boldsymbol{k}\right|}=\sqrt{1-\beta^{2}\gamma^{2}\omega_{p}^{2}/\boldsymbol{k}^{2}},\label{eq:theta2}
\end{equation}
which is neither longitudinal nor transverse when $\omega_{p}\ne0.$
\item Current-plasmons in the lab frame. Finally, the dispersion relations
for the current plasmons in the lab frame $x$ is 
\begin{equation}
\omega-\beta_{j}k^{j}=0,
\end{equation}
whose manifestly covariant form is
\begin{equation}
k^{\mu}u_{0\mu}=0.
\end{equation}
Note that the current-plasmon is not simply the magnetostatic current
mode in the rest frame Doppler-shifted into the lab frame $x$. In
the lab frame $x$, the mode carries a time-dependent electrical component
that does not exist in the rest frame. To see that this is an oblique,
propagating electromagnetic excitation, we take the example of 
\begin{align}
\psi^{\prime} & =\left\{ \left(0,0,0,\frac{\omega_{p}^{2}}{K^{2}}\right),\left(0,0,0,1\right),0\right\} ,\\
k^{\prime} & =(0,K,0,0),
\end{align}
which is a static current-plasmon in the rest frame $x^{\prime}$
with wavenumber in the 1st-direction and magnetically polarized in
the $2$nd-direction. Assuming the plasma flow in the lab frame $x$
is in the 1st- and 3rd-directions, i.e., 
\begin{equation}
\Lambda^{-1}=\Lambda\left(-\beta_{1},0,-\beta_{3}\right),\label{eq:Lamda3-1}
\end{equation}
then in the lab frame $x$, we find 
\begin{equation}
\delta\boldsymbol{E}=\frac{\omega_{p}^{2}}{K\gamma}(-i\beta_{3},0,\beta_{1})^{T}.
\end{equation}
The cosine of the pitch angle $\theta_{3}$ between $\delta\boldsymbol{E}$
and $\boldsymbol{k}$ is 
\begin{equation}
\cos\theta_{3}\equiv\frac{\left|\delta\boldsymbol{E}\cdot\boldsymbol{k}\right|}{\left|\delta\boldsymbol{E}\right|\left|\boldsymbol{k}\right|}=\sqrt{\frac{\beta_{3}^{2}}{\left(\gamma^{2}-1\right)\left(1-\beta_{3}^{2}\right)}},\label{eq:theta3}
\end{equation}
which is neither longitudinal nor transverse in general. 
\end{enumerate}
As a potential application of the theory developed, the polarization
angles of the oblique photons, plasmons, and current-plasmons given
by Eqs.\,(\ref{eq:theta1}), (\ref{eq:theta2}) and (\ref{eq:theta3})
can be explored as diagnostic measures of density and/or flow velocity
of relativistic plasmas. 

We now focus on the photons in plasmas. For this purpose, it is necessary
to consider plasmons as well because photons and plasmons resonate
at $(\omega,\boldsymbol{k})=(\gamma\omega_{p},\gamma\boldsymbol{\beta}\omega_{p}).$
On the other hand, photons do not resonate with current-plasmons,
which can thus be neglected in this context. We just proved that both
photons and plasmons are oblique and electromagnetic in general. How
do we tell them apart then? Is there a Lorentz invariant label that
can distinguish photons from plasmons? It turns out that covariant
compressibility defined by
\begin{equation}
\chi\equiv\delta u_{\mu}k^{\mu}
\end{equation}
is such a label. From the eigenvector spaces listed above, we can
calculate that for photons $\chi=0$ and for plasmons $\chi=\omega_{p}.$
In this sense, plasmons are covariantly compressible electromagnetic
excitations, and photons (and current-plasmons) are covariantly incompressible
ones. This is a more precise and more general description than the
conventional picture of plasmons being electrostatic and longitudinal
while photons are electromagnetic and transverse. 

\begin{figure}[ht]
\centering \includegraphics[width=5cm]{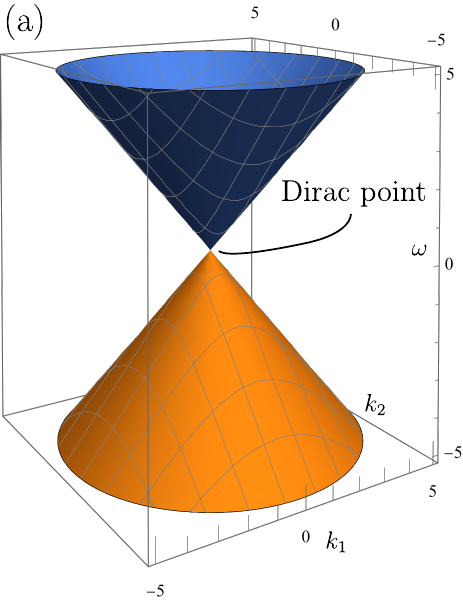} ~~~~\includegraphics[width=5cm]{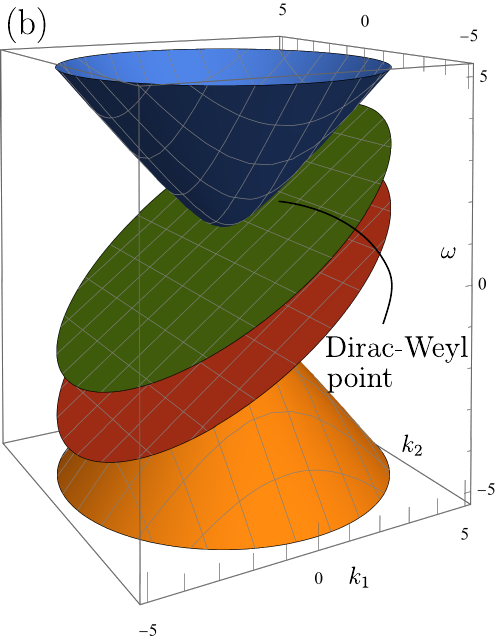}
\caption{Photon topology in vacuum (a) and in a relativistic plasma with $\beta=0.6$
(b). The dispersion relations of photons, anti-photons, plasmons,
and anti-plasmons are plotted in blue, orange, green, and red, respectively.
The momentum space is shown as 2D. Plotted $k_{1}$, $k_{2}$, and
$\omega$ are normalized by $\omega_{p}$. The photon topology in
vacuum (a) is singular at the $(\omega,\boldsymbol{k})=(0,\boldsymbol{0})$
Dirac point of the light cone, which is the fundamental reason that
photons do not admit a spin angular momentum operator \citep{Palmerduca2023}.
Plasmas remove the photon's Dirac point by giving it an effective
mass of $\omega_{p}$, but create a tilted Dirac-Weyl point at $(\omega,\boldsymbol{k})=(\gamma\omega_{p},\gamma\beta\omega_{p},0)$
by reviving the virtual longitudinal photon (b). }
\label{pt}
\end{figure}

One of the motivations of the present study is to explore the photon
and plasmon topology, which is intimately related to topological waves
in continuous media \citep{Delplace2017,Parker2020,Gao2016,Yang2016,Parker2020a,Faure2023,fu2021topological,Fu2022a,Delplace2022,Qin2023,Zhu2023,Fu2023a,Fu2023,WangB2016,Bernety2023},
spin angular moment of photons or the non-existence thereof \citep{Akhiezer1965,Enk1994,Bliokh2015,BialynickiBirula2011,Leader2013,Leader2016,Leader2019,Palmerduca2023},
and the non-Hermitian or PT-symmetric classical physics \citep{Bender1998,Bender2007,Qin2019KH,Fu2020KH,Zhang2020PT,Qin2021,Israeli2023a,Meng2023}.
In particular, the comparison between photon topology in vacuum and
in plasmas is revealing. It has now been established that the angular
momentum for vacuum photons can't be split into orbital angular momentum
and spin angular momentum parts \citep{Enk1994,BialynickiBirula2011,Leader2013,Leader2016,Leader2019}.
Unlike massive particles, photons in vacuum do not admit a spin angular
momentum operator. This fact is fundamentally attributed to the topological
singularity at the $(\omega,\boldsymbol{k})=(0,\boldsymbol{0})$ Dirac
point of the light cone \citep{Palmerduca2023}, see Fig.\,\ref{pt}(a).
Is the photon topology nontrivial in plasmas as in vacuum? The dispersion
relations of photons in plasma is shown in Fig.\,\ref{pt}(b) along
with those of plasmons. Observe that the $(\omega,\boldsymbol{k})=(0,\boldsymbol{0})$
Dirac point for photons in vacuum does exist in plasmas anymore because
of the gap of $2\omega_{p}$ between photons and anti-photons, which
is evident from the dispersion relation given by Eq.\,(\ref{eq:photonDR}).
This can be viewed as photons gaining an effective mass of $\omega_{p}$
from the plasma. However, a new exceptional point at $(\omega,\boldsymbol{k})=(\gamma\omega_{p},\gamma\beta\omega_{p},0)$
emerges due to the interaction between photons and plasmons. At this
exceptional point, the 2D photon eigenmode bundle and the 1D plasma
eigenmode bundle are not distinguishable. For this reason, it should
be called a Dirac-Weyl point. It should be removed from the base manifold
of the 2D photon eigenmode bundle, which renders the base manifold
of photon eigenmode bundle non-contractible, akin to the situation
of photons in vacuum \citep{Palmerduca2023,Qin2023}, and the associated
vector bundle topology is expected to be nontrivial and interesting.
But the topology will be different from the photon topology in vacuum
\citep{Palmerduca2023}, because the exceptional point at $(\omega,\boldsymbol{k})=(\gamma\omega_{p},\gamma\beta\omega_{p},0)$
is a tilted Dirac-Weyl point \citep{Qin2023} instead of an upright
Dirac point. Plasmas remove the photon's upright Dirac point by giving
it effective mass, but create a tilted Dirac-Weyl point by reviving
the virtual longitudinal photon. These topics and the implications
on the angular momentum of photons in plasmas will be reported in
future publications.
\begin{acknowledgments}
This research was supported by the U.S. Department of Energy (DE-AC02-09CH11466). 
\end{acknowledgments}

\bibliographystyle{apsrev4-2}
\bibliography{../Refs/Refs}

\end{document}